\documentclass[9pt]{article}

\usepackage{cite}
\usepackage{amsmath,amssymb,amsfonts}
\usepackage{algorithmic}
\usepackage{graphicx}
\usepackage{textcomp}
\usepackage{xcolor}
\usepackage{spconf}
\usepackage{hyperref}
\usepackage{url}
\usepackage{makecell}
\usepackage{comment}
\usepackage{multirow}
\usepackage[normalem]{ulem}
\usepackage{subcaption}

\newif\ifcomments
\commentsfalse

\ifcomments

\newcommand{\TODO}[1]{\textcolor{red}{{#1}}}
\newcommand{\BT}[1]{\textcolor{blue}{{#1}}}
\newcommand{\AC}[1]{\textcolor{violet}{{#1}}}

\else

\newcommand{\TODO}[1]{}
\newcommand{\BT}[1]{}
\newcommand{\AC}[1]{}

\fi

\title{OF-S\MakeLowercase{em}W\MakeLowercase{at}: High-Payload Text Embedding for Semantic Watermarking of AI-Generated Images with Arbitrary Size	\vspace{-0.2cm}}
\name{Benedetta Tondi, Andrea Costanzo, Mauro Barni
	\vspace{-0.3cm}
}
\address{University of Siena, Siena, Italy}
\begin{document}
%
\maketitle
\begin{abstract}
We propose a high-payload image watermarking method for textual embedding,
where a
semantic description of the image - which may also correspond to the input text prompt-, 
 is embedded inside the image.
In order to be able to robustly embed high payloads in large-scale 
 images - such as those produced by modern AI generators - 
 the proposed approach 
 %
 builds upon a traditional watermarking scheme that exploits orthogonal and turbo codes for improved robustness, and  integrates frequency-domain embedding and perceptual masking techniques to enhance watermark imperceptibility.
Experiments show that the proposed method 
is extremely robust against a wide variety of image processing, and the embedded text can be retrieved also after traditional and AI inpainting, permitting to unveil the semantic modification the image has undergone via image-text mismatch analysis.
%
\end{abstract}

 \TODO{
 From the current title ChatGPT understands perfectly what we are proposing. So I would opt for this. 
 An alternative: "OF-SemWat: Robust Embedding of Semantic Text in AI-Generated Images of Arbitrary Resolution for Proactive Detection "\\ 
}



\section{Introduction}
\label{sec:intro}

Modern AI models offer incredible benefits by enabling anyone to easily create extremely realistic images. Nonetheless, they also pose significant risks related to image misuse, including the spread of misinformation and potential public harm.
Detecting AI-generated images has then become crucial  for mitigating these 
risks. 
%
In addition to passive methods - which seek to detect subtle artifacts in AI-generated images, distinguishing them from real ones or 
attributing them to the generative AI model \cite{tariang2024synthetic} -, proactive watermarking techniques, which embed imperceptible signals into AI-generated images to facilitate their detection/attribution, have recently attracted growing interest from both policymakers and industry stakeholders \cite{jiang2024watermark}.
In 2021, the Coalition for Content Provenance and Authenticity, or C2PA ({\footnotesize\url{https://c2pa.org}}), was launched by major companies 
to develop an open technical standard for tracking the provenance and authenticity of digital media.
Based on the stage at which the watermark is embedded into an AI-generated image, methods are categorized 
as pre-generation \cite{wen2023tree}, post-generation \cite{tancik2020stegastamp, HiDDeN}, and in-generation \cite{Fei2022, fernandez2023stable,fei2024wide,fei2023robust,kim2024wouaf}. 
In particular, in-generation methods modify the parameters of the generative model during training, ensuring  that the output images contain a watermark,
while in post-generation watermarking instead the image is watermarked in a later stage, after the model has been trained.
By performing
joint model and image watermarking, in-generation methods are typically adopted 
to protect the ownership of AI models \cite{Li2021b}.
For watermark-based authentication, i.e. when the goal is to proactively embeds watermarks in AI-generated images for detection/attribution purposes, there is no need to watermark the model itself and post-generation methods are often used.

A common shortcoming of post-generation watermarking methods developed so far, is that they can only work with small (or medium) size image, typically 128$\times$128 or 256$\times$256 at most \cite{tancik2020stegastamp, HiDDeN,jia2021mbrs,ma2022towards}, with very few methods reaching 512$\times$512 \cite{zhang2024editguard}, embedding low payloads.
However, images generated by modern AI models 
are often large in size (1024×1024 pixels or higher).
An approach capable to work with images of arbitrary resolutions has been developed by the Adobe company \cite{bui2023trustmark}.
However, being the method based on resolution downsampling to 256x256,  
the payload that can be embedded remains limited (up to 150/200 bit approx).


In this paper, we propose a post-generation old fashioned (OF) watemarking method, 
for embedding arbitrary size - and in particuar large scale - images with high-payload watermarks. The method is employed within a semantic watermarking framework, called OF-SemWat, where either a textual description of the image - or the input prompt itself - is embedded into the image.
Beyond its potential application for establishing ownership, 
the semantic watermark provides information about the original content of the image,
helping to determine {\em whether} and {\em how} the image has been semantically altered (e.g., through the use of a different AI or an AI from a non-trusted provider), 
through an analysis of the mismatch between the image and the recovered text. The semantic watermark can also serve as a useful means for  
characterizing the intent behind image manipulations.
%
%
To achieve our goal of embedding large-scale images with high-payload watermarks, 
we resort to a traditional robust watermarking technique exploiting orthogonal codes and turbo codes \cite{Abrardo2005}.
In order to improve the imperceptibility of the watermark, we resort to frequency-domain embedding and integrate perceptual masking techniques.
The method is a very flexible one which can work with images of arbitrary size.
%
Experiments 
carried out with AI-generated images from modern generators (SD-XL, Flux and Mystic), and also real images taken from cameras
 \footnote{Likewise any image watermarking method, the proposed method is not just for AI-generated images and can also be applied to real images.
 	},
show that the proposed method is very robust to a wide variety of processing, namely, JPEG compression, noise addition and geometric modification (resizing, rotation, cropping),  and the embedded textual description can be retrieved also when the semantic content of the image is modified via traditional or AI inpainting. 
Moreover, from our experiments we found that, when the bit error rate is low (lower than 12\% approx), 
the use of an LLM for text error correction enables the textual description to be recovered either exactly or without semantic alteration.

\vspace{-0.3cm}
\section{Proposed watermarking  system}
\label{sec:method}


The proposed watermarking approach extends the strongly robust traditional watermarking method in \cite{Abrardo2005}   by integrating several perceptual techniques to enhance watermark invisibility (see Section  \ref{sec:OFBmethod}).
Secrecy of the watermark is guaranteed by the use a secret key, shared by encoder and decoder.
In contrast with current network-based image watermarking systems, where robustness is  typically achieved only against the types of processings considered during training,  the proposed method achieves 
robustness against  
{\em any} kind of
image modification, entailing both processing operations, e.g., JPEG compression and noise addition, and semantic modification, e.g., inpainting.
%
The method is very flexibile, capable to work with any input size and payload, that can be made very high for large-size images.
%
%
%

\begin{figure*}
	\centering\includegraphics[width=0.97\textwidth]{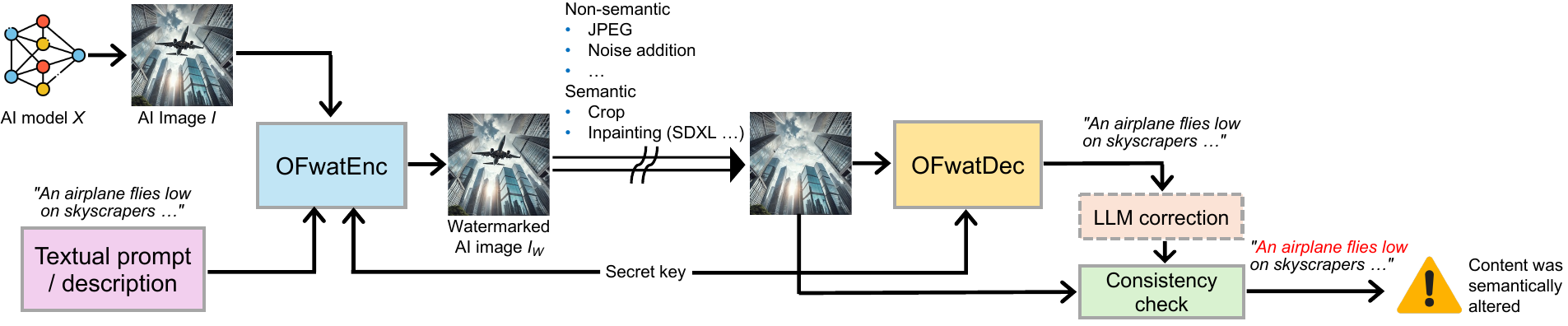}
	\caption{Illustration of the proposed OF-SemWat for semantic watermarking of AI images of arbitrary size, and especially large-scale images (1024$\times 1024$ and above).
		Besides potentially serving as a means of establishing ownership, the semantic watermark helps reveal content modifications the image may have undergone.
	}
	\vspace{-0.4cm}
	\label{fig.scheme}
\end{figure*}

The method is employed within the  
proposed OF-SemWat framework 
illustrated in Fig.  \ref{fig.scheme}, for semantic watermarking of AI-generated images with arbitrary size, providing a useful tool for semantic manipulation detection.
A textual description of the image is binary encoded 
and embedded inside the AI-generated image by the encoder (OFwatEnc).
Once the watermark is extracted from the decoder (OFwatDec), image-text consistency analysis is performed to decide whether the image has been semantically altered or not, and how. 
We also (optionally) resort to an LLM (GPT4)  for text error correction, which allows us to retrieve the original textual description/prompt, or a text with exactly the same semantic meaning, when a few 
bit errors occur in the retrieved bit string
affecting the textual description.
%
Note that, in principle, depending on the application scenario, any description useful for accompanying the image can be embedded via the proposed watermarking method. 
Finally, we point that ownership can also be established based on the recovered text, e.g., by detecting whether it is a random sequences of characters or a meaningful and structured text, via a dedicated system,
or reserving CRC (Cyclic Redundancy Check) bits to be used for watermark authentication \cite{shih2017digital}.

\begin{figure}[h!]
  \centering
  \begin{subfigure}{\columnwidth}
    \centering
    \includegraphics[width=0.3\textwidth]{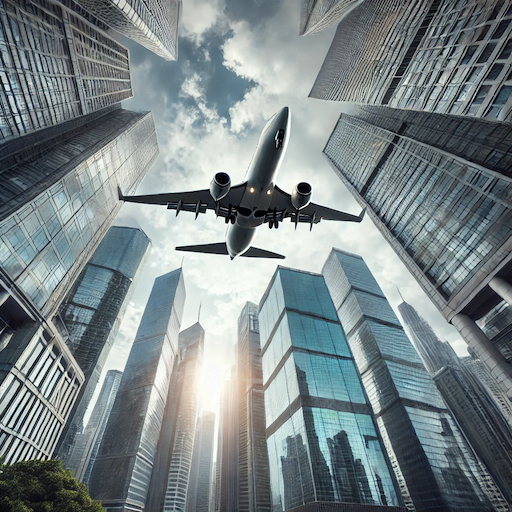}
    \includegraphics[width=0.3\textwidth]{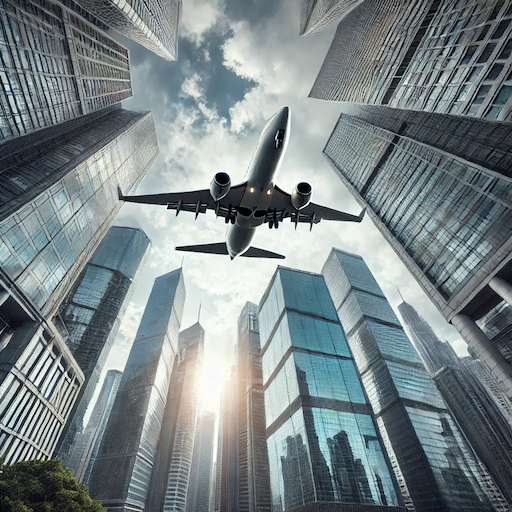}
    \includegraphics[width=0.3\textwidth]{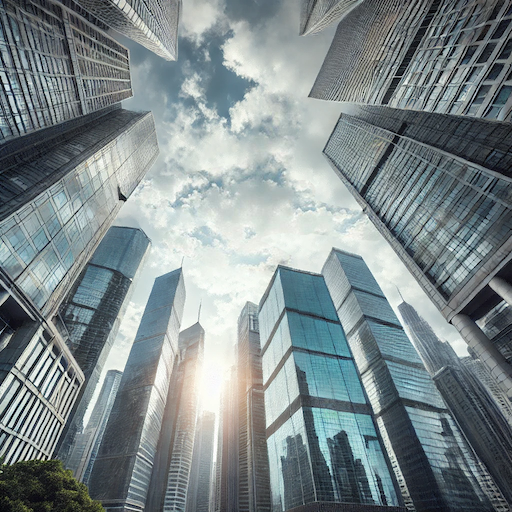}
    \caption*{\em{An \textcolor{red}{airplane flies low} near modern skyscrapers in a dense city. The glass buildings reflect sunlight, and the sky is partly cloudy. The perspective emphasizes the dramatic proximity of the \textcolor{red}{plane} and towers.}}
  \end{subfigure}

  \vspace{0.1cm} 

  \begin{subfigure}{\columnwidth}
    \centering
    \includegraphics[width=0.3\textwidth]{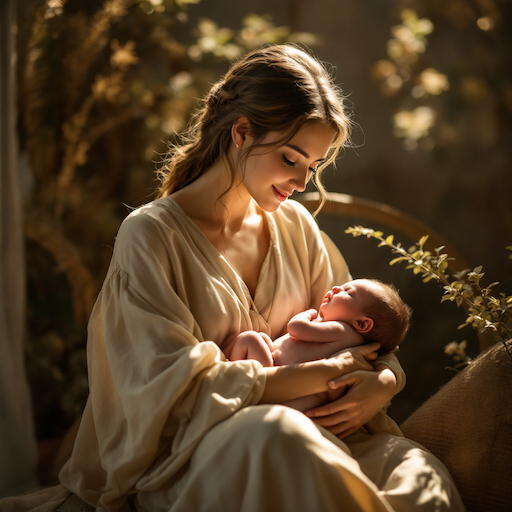}
    \includegraphics[width=0.3\textwidth]{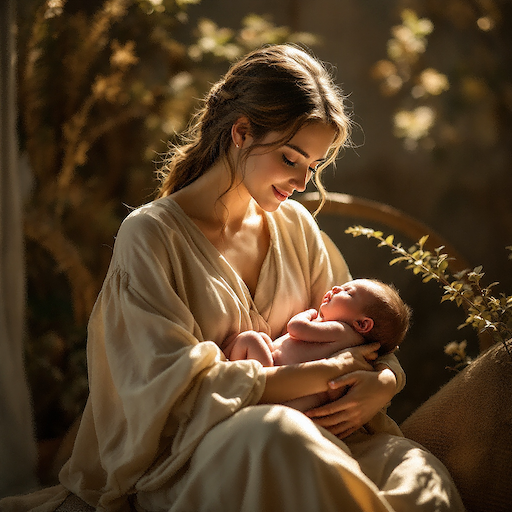}
    \includegraphics[width=0.3\textwidth]{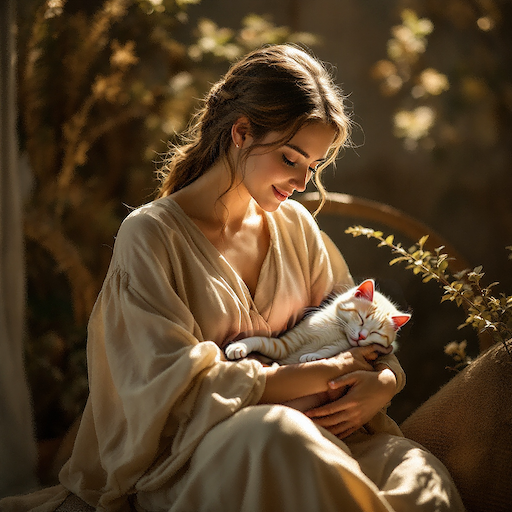}
    \caption*{\em{A serene young woman in a flowing beige gown, sitting and cradling a \textcolor{red}{newborn baby} in her lap. Warm sunlight filters through, casting a soft glow, evoking classic Madonna and \textcolor{red}{Child} iconography.}}
  \end{subfigure}

  \caption{Examples of 
  watermarked images. From left to right: original image, watermarked image,  AI-edited watermarked image (using online Freepik Pikaso). Original and recovered text are equal (bit error rate is 0). Other examples can be found 
at  {\footnotesize 
\url{https://tinyurl.com/OFSemWatExample}}.}
\vspace{-0.5cm}
  \label{fig:example}
\end{figure}

Figure \ref{fig:example} shows some examples of watermarked image, along with the embedded watermark description, and the corresponding watermark extracted from the OFWatDec after semantic modification via AI inpainting.

\subsection{Robust and flexible image watermarking}
\label{sec:OFBmethod}

%
%
As mentioned,  the proposed approach builds upon the robust watermarking method in \cite{Abrardo2005}, by integrating several perceptual techniques to enhance watermark invisibility.
Specifically, the $l$-bit watermark message is embedded in the Discrete Fourier Transform (DFT) domain, where medium frequency coefficients are selected from the amplitude spectrum of the DFT applied to the luminance channel $Y$ in the YCbCr color space of the input image $I$. 
While modifications at low frequencies are too visible, and high frequencies are often degraded by image processing,
medium-frequency bands offer a good balance between imperceptibility and robustness.
To further improve perceptual quality, each host coefficient is weighted, prior to embedding, according to a perceptual sensitivity mask proposed by Watson \cite{Watson1993}, which models the human visual system’s  sensitivity to spatial frequencies. As to the watermark, the 
 message bits 
 are first turbo-encoded and then embedded using the Quasi-Orthogonal Dirty Paper Coding (QO-DPC) scheme with fixed distortion \cite{Abrardo2005}, 
 exploiting orthogonal codes where, similarly to spread spectrum, each bit is spread across $s$ distinct host coefficients, randomly shuffled based on the secret key to improve both robustness and security (we refer to \cite{Abrardo2005} for the details).
The modified coefficients are then restored to their original positions within the DFT domain, and the inverse DFT is applied to reconstruct the preliminary watermarked luminance channel $Y_w'$. 
To further improve imperceptibility, a variance-based perceptual mask $M$ is computed from $Y_w'$, 
and the embedding strength is adaptively modulated, enforcing a stronger watermark in high-variance regions and a weaker one in smoother areas, exploiting the reduced sensitivity of the human visual system to distortions in regions of high local variance.
The final watermarked channel $Y_w$ is obtained as a pixel-wise linear combination of the original cover image $Y$ and $Y_w'$, weighted by the variance mask $M$:
$Y_w = (1 - M) \cdot Y + M \cdot Y_w'$. Finally, $Y_w$ is converted back to the RGB color space to produce the final watermarked image $I_w$.
The watermark is extracted by OFwatDec by first extracting the coefficients in the positions indicated by the secret key and then choosing the codeword (uniquely mapping to the $l$-bit string) via correlation-based decoding (see \cite{Abrardo2005}).

%
An implementation of our method is made publicly available at  	{\footnotesize 
\url{https://tinyurl.com/OFSemWatCode}
}.

\section{Experimental analysis}
\label{sec:experiments}

\subsection{Methodology}

\textbf{Dataset.}
To assess the performance of the proposed system, we considered 100 images of general content generated from Stable Diffusion (SD)-XL, FLUX.1-dev 
and Mystic (Freepik), 
which have original size 1024$\times$1024 (SD and Flux) and 2048x2048 (Mystic). Mystic images are resized to 1024$\times$1024. 
 %
We also run some tests with 100 real, large images coming from the RAISE1k dataset \cite{dang2015raise}, by downsampling them to a size 1256$\times$834 (same no. of pixel of $1024\times 1024$)  and 2512$\times$1668, which keep the original aspect ratio. 

With regard to the non semantic modifications, 
the following post-processings are considered: JPEG compression, resizing (down and up), rotation, crop (central cropping of a portion of the image) and Gaussian noise addition,  implemented through the Python PIL library.
Semantic modifications are performed automatically to the images via: i) traditional splicing; ii) AI-based inpainting. In i), the image is first segmented using SegmentAnything \cite{kirillov2023segany}; then, segments of the watermarked image are replaced with the corresponding non-watermarked image. 
Given a desired splicing percentage $p$, linear programming is used to get the bet pool of to-be-replaced segments summing up to $p$.
For ii), SD-XL inpainting is used, guided by a binary mask (as in \cite{zhang2024editguard}), 
which is obtained by segmenting the image with SegmentAnything and selecting the portions as described above.  
$\\$
\textbf{Watermark settings.}
For every image, a description of 200 characters - raised to 600 for the tests with higher payloads-,  is obtained through GPT4, which serves as watermark. The text is converted to a bit string using 5 bit per each character (for all the alphabet letters, blank character and main punctuation), resulting in a message payload $l$ of 1000 bit - raised to 3000 bit. 
When not specified otherwise, the spreading factor $s$ is set to $128$ (default).  
$\\$
\textbf{Comparison with the state-of-the-art.}
As mentioned in the introduction, existing post-generative methods either do not work with large-scale images or can not embed large payloads, hence 
they cannot be considered for comparison. 
A method which is similar in spirit to our method is EditGuard \cite{zhang2024editguard}, where a proactive method with tamper localization capabilities is proposed. The method works by embedding a robust watermark of 64 bit and a localization  semi-fragile watermark (i.e.  a given image) which is at the same time robust against processing
and fragile to tampering. 
The tampering is localized performing the difference between the localization watemark and decoded watermark. The input size for this network is 512$\times$512 (for non-square images, a square crop of the maximum dimension, i.e., the smallest between width and height, is taken before resizing). 
%
A drawback of the method in \cite{zhang2024editguard}, also claimed in later works and confirmed by our experiments, is that robustness is achieved only against the specific processings considered during training. 
On the contrary, not relying on learning-based processes, our method can get robustness against any kind of manipulation.


\subsection{Results}

Table 	\ref{table:resultsAIimages}  reports the accuracy of watermark extraction, namely the bit error rate (BER), on both clean and processed images, achieved by our method and EditGuard on AI images. The average PSNR and SSIM on clean images is respectively $34.3$ dB and $0.92$ for OF-SemWat and $37.5$ dB and $0.98$ for EditGuard \cite{zhang2024editguard}. For OF-SemWat, the payload  is set to 1000 (200 characters description). 
The performance of localizazion of EditGuard are also reported, in terms of  average localization accuracy (LocAcc),  and the Fscore.
EditGuard achieves good performance on AI-inpaining and Gaussian Noise addition (with small $\sigma$), 
which correspond to processings 
(and strenghts) considered during training. In the other cases, robustness is poor. Even if it is reported in all the cases, Fscore metrics is only meaningful in the case of local tamperings (a NaN is returned in the other cases, indicated with -  in the tables).  LocAcc = 0 indicates that all the bits are wrongly classified (a completely white mask  - all bit detected as tampered - in the case of black groud truth).
OF-SemWat achieves robustness against all the processing, and the BER is always 0 or less than 1\% in almost all cases, and it becomes large only when the  processing is applied with very large strenght (e.g., JPEG with QF = 5 or noise with $\sigma = 20$), 
which however results 
in a severe degradation of the image quality.

%

\enlargethispage{\baselineskip}

\begin{table}[htbp]
	\scriptsize
	\setlength{\tabcolsep}{1.9pt}
	\centering
	\caption{Watermark performance (BER \%) on 
		AI-images - clean and processed. 
		The payload for OF-SemWat is  set to $l =1000$ bit (while \cite{zhang2024editguard}  uses $64$ bit). \vspace{-0.2cm}
	}
\begin{tabular}{l | c | c | c c c | c c c | c c | c c}
	\hline
	& & Clean 
	& \multicolumn{3}{c|}{Crop (\%)} 
	& \multicolumn{3}{c|}{JPEG (QF)} 
	& \multicolumn{2}{c|}{Resize} 
	& \multicolumn{2}{c}{Rotation} \\  
	\Xhline{2\arrayrulewidth}
	& & & 10 & 20 & 30 
	& 70 & 30 & 5  
	& 0.7 & 1.7 
	& 5\textdegree & 10\textdegree \\  
	\Xhline{2\arrayrulewidth}
	OF-Sem & BER & 0 & 0 & 0.2 & 4.7 & 0 & 0.2 & 49.9 & 0 & 1.9 & 0.1 & 0.3 \\ \hline
	\multirow{3}{*}{\cite{zhang2024editguard}} 
	& BER     & 0 & 49.2 & 50.0 & 47.9 & 16.1 & 28.6 & 43.9 & 68.5 & 20.0 & 48.9 & 50.4 \\  
	\cline{4-13}
	& LocAcc  &   & 0 & 0 & 0 & 0 & 0 & 0 & 0 & 0 & 0 & 0 \\    
	\cline{4-13}
	& Fscore  &   & - & - & - & - & - & - & - & - & - & - \\  
	\hline
	\multicolumn{3}{c|}{} 
	& \multicolumn{2}{c|}{Noise ($\sigma$)} 
	& \multicolumn{3}{c|}{Splicing (\%)} 
	& \multicolumn{3}{c|}{AI-inpaint (\%)} 
	& & \\  
	\cline{4-11}
	\multicolumn{3}{c|}{} 
	& 5 &  \multicolumn{1}{c|}{20}
	&\multicolumn{1}{c}{10} & 15 & \multicolumn{1}{c|}{25} 
	& \multicolumn{1}{c}{5} & 10 & 20 
	& & \\  
	\cline{1-11}
	OF-Sem  & BER & &  \multicolumn{1}{c}{0} & \multicolumn{1}{c|}{0}   & \multicolumn{1}{c}{0} & 0 & \multicolumn{1}{c|}{0}   &  \multicolumn{1}{c}{0} & 0.2 & 3.1 & & \\  
	\cline{1-11}
	\multirow{3}{*}{\cite{zhang2024editguard}} 
	& BER & & 0 & \multicolumn{1}{c|}{17.6}   & \multicolumn{1}{c}{0} & 0 & \multicolumn{1}{c|}{0}   & \multicolumn{1}{c}{0}  & 0 & 0 &  & \\  
	\cline{4-11} 
	& LocAcc & & 78.9 &  \multicolumn{1}{c|}{2.4}   & \multicolumn{1}{c}{99.7} & 98.0 & \multicolumn{1}{c|}{95.9}   & \multicolumn{1}{c}{99.1}  & 98.4 & 96.4 & & \\  
	\cline{4-11} 
	& Fscore & & - &  \multicolumn{1}{c|}{-}   & \multicolumn{1}{c}{93.8} & 93.6 & \multicolumn{1}{c|}{92.2}   & \multicolumn{1}{c}{91.6}  & 92.7 & 91.7 & & \\  
	\cline{1-11}
\end{tabular}
	\label{table:resultsAIimages}
\end{table}

As expected, we got similar results when the watermark is embedded on RAISE images.  Table  \ref{table:resultsRAISEsmall} reports the performance 
for the 1256$\times$834 size, and same payload $l = 1000$ (200 characters). In this case, the average PSNR and SSIM on clean images is respectively $34.6$ dB and $0.95$ for OF-SemWat and $35.7$ dB and $0.97$ for EditGuard.
%
\begin{table}[htbp]
	\scriptsize
	\setlength{\tabcolsep}{1.9pt}
	\centering
	\caption{Watermark performance (BER \%) on 
		AI-images - clean and processed. 
		The payload for OF-SemWat is  set to $l =1000$ bit (while \cite{zhang2024editguard}  uses $64$ bit).
		\vspace{-0.2cm}
	}
	\begin{tabular}{l | c | c | c c c | c c c | c c | c c}
		\hline
		& & Clean 
		& \multicolumn{3}{c|}{Crop (\%)} 
		& \multicolumn{3}{c|}{JPEG (QF)} 
		& \multicolumn{2}{c|}{Resize} 
		& \multicolumn{2}{c}{Rotation} \\  
		\Xhline{2\arrayrulewidth}
		& & & 10 & 20 & 30 
		& 70 & 30 & 5  
		& 0.7 & 1.7 
		& 5\textdegree & 10\textdegree \\  
		\Xhline{2\arrayrulewidth}
		OF-Sem & BER & 0 & 0 & 2.9 & 10.8 & 0 & 2.7 & 50.1 & 0.2 & 7.8 & 1.6 & 2.7 \\ \hline
		\multirow{3}{*}{\cite{zhang2024editguard}} 
		& BER     & 0 & 49.7 & 50.1 & 48.9 & 17.4 & 28.0 & 41.9 & 64.8 & 19.8 & 49.6 & 53.1 \\  
		\cline{4-13}
		& LocAcc  &   & 0 & 0 & 0 & 0 & 0 & 0 & 0 & 0 & 0 & 0 \\    
		\cline{4-13}
		& Fscore  &   & - & - & - & - & - & - & - & - & - & - \\  
		\hline
		\multicolumn{3}{c|}{} 
		& \multicolumn{2}{c|}{Noise ($\sigma$)} 
		& \multicolumn{3}{c|}{Splicing (\%)} 
		& \multicolumn{3}{c|}{AI-inpaint (\%)} 
		& & \\  
		\cline{4-11}
		\multicolumn{3}{c|}{} 
		& 5 &  \multicolumn{1}{c|}{20}
		&\multicolumn{1}{c}{10} & 15 & \multicolumn{1}{c|}{25} 
		& \multicolumn{1}{c}{5} & 10 & 20 
		& & \\  
		\cline{1-11}
		OF-Sem  & BER & &  \multicolumn{1}{c}{0} & \multicolumn{1}{c|}{0}   & \multicolumn{1}{c}{0} & 0 & \multicolumn{1}{c|}{0}   &  \multicolumn{1}{c}{1.5} & 4.2 & 12.3 & & \\  
		\cline{1-11}
		\multirow{3}{*}{\cite{zhang2024editguard}} 
		& BER & & 0 & \multicolumn{1}{c|}{17.8}   & \multicolumn{1}{c}{0} & 0 & \multicolumn{1}{c|}{0}   & \multicolumn{1}{c}{0}  & 0 & 0 &  & \\  
		\cline{4-11} 
		& LocAcc & & 84.6 &  \multicolumn{1}{c|}{3.6}   & \multicolumn{1}{c}{99.3} & 98.2 & \multicolumn{1}{c|}{96.5}   & \multicolumn{1}{c}{99.1}  & 98.5 & 97.1 & & \\  
		\cline{4-11} 
		& Fscore & & - &  \multicolumn{1}{c|}{-}   & \multicolumn{1}{c}{91.3} & 93.6 & \multicolumn{1}{c|}{93.1}   & \multicolumn{1}{c}{90.2}  & 91.5 & 92.5 & & \\  
		\cline{1-11}
	\end{tabular}
	\label{table:resultsRAISEsmall}
\end{table}
\begin{table}[htbp]
	\footnotesize
	\setlength{\tabcolsep}{1.9pt}
	\centering
	\caption{ 
	Watermark performance  (BER (\%)) on very large-scale images from RAISE  ($2512\times1668$), for two different values of the spreading factor  $s$. The payload is $l = 3000$ bit when $s=64$ and $l = 1500$ bit when $s=128$.
		\vspace{-0.2cm}
	}
	\begin{tabular}{ l |c c  c |  c c c | c c  c c  c c}
	%
	\hline
	&  \multicolumn{3}{c|}{Crop (\%)} &  \multicolumn{3}{c|}{JPEG (QF)} & 
	\multicolumn{2}{c|}{Resize} & 
	\multicolumn{2}{c|}{Rotation} & \multicolumn{2}{c}{Noise ($\sigma$)} \\  \Xhline{2\arrayrulewidth}
		&  10 & 20 & 30  &  70 &30& 5 & 
	0.7 & \multicolumn{1}{c|}{1.7} & 
5 \textdegree & \multicolumn{1}{c|}{10 \textdegree} & 5 & 20\\ 
	\hline
s = 128	& 0.9 & 13.9 & 34.8   & 0 & 0  & 19.0  & 0 &  \multicolumn{1}{c|}{0}  & 0.1   & \multicolumn{1}{c|}{0.2} & 0 & 0 \\    	  \hline
s = 64 & 0  & 3.6 & 23.4   & 0 & 0  & 4.2  & 0 &  \multicolumn{1}{c|}{0} & 0   & \multicolumn{1}{c|}{0} & 0 & 0\\   \hline
		 
		 &  \multicolumn{3}{c|}{Splicing (\%)} & 
	\multicolumn{3}{c|}{AI-inpaint (\%)}   &  
	& & & & &  \\  
	 \cline{1-7}
	&  10 & 15 & 25 & 
	5 & 10 & 20  &  && & & &    \\  
	\cline{1-7}
s = 128	 & 0  & 0 & 0.1   & 0.1 & 0.5  & 11.6  &  &  & & & & \\   	
	\cline{1-7}
s = 64  & 0  & 0 & 0   & 0 & 0.3  & 9.9  &  & &   & & & \\ 	  
\cline{1-7}
\end{tabular}
	\label{table:resultsRAISE}
\end{table}
\begin{table}
		\footnotesize
	\setlength{\tabcolsep}{2pt}
	\centering
	\caption{Semantic similarity between the recovered text before and after the error correction performed by the LLM.
	\vspace{-0.2cm}
	}
	\begin{tabular}{c | c  c | c c | c c | c c |c   c | }
		
		\hline
		\text{BER} &\multicolumn{2}{c|}{15\%} &\multicolumn{2}{c|}{12\%}& \multicolumn{2}{c|}{10\%}  &  \multicolumn{2}{c|}{5\%}  & \multicolumn{2}{c|}{3\%}
		  \\ \hline
		  		  & bef. & aft. & bef. & aft. &bef. & aft. & bef. & aft. & bef.  & aft. \\ \cline{2-11}
		 SemSim \cite{reimers-2019-sentence-bert} & 0.25 & 0.58 &0.38& 0.86 & 0.44 & 0.95  & 0.66 & 0.96 & 0.83 & 0.99 \\ \hline
	\end{tabular}
	\label{table:resultsAIcorrection}
	\vspace{-0.2cm}
\end{table}
The results obtained in the case of very large-scale images ($2512\times1668$), for a payload of 3000 bit (600 charatecters description), setting $s$ to $64$, are reported in Table 	\ref{table:resultsRAISE}.
The PSNR/SSIM in this case is $31.5$ dB / $0.92$. 
The table also shows the results obtained by reducing $l$ to $1500$ (300 characters) and increasing $s$ to 128,
to exchange the payload with robustness (the length of the encoded bit sequence is kept fixed, i.e., equal to 192000). 
As expected, robustness improves in the second case.

Notably, using an LLM for text error correction, a certain number of bit errors can be corrected a posteriori, enabling  recovery
of the exact original text or a text with only minor differences, that preserve the same semantic meaning.
%
%
Table \ref{table:resultsAIcorrection} shows the results we got 
using the sentence-transformer model all-MiniLM-L6-2, based on Semantic-BERT \cite{reimers-2019-sentence-bert},
to measure sentence similarity,
as the cosine distance between the two sentence embeddings. When BER is up to 12\%, 
the similarity score is around 0.9, in which case the semantic meaning of the sentence can be fully recovered. Thus, the semantic modification can be revealed.
Some examples of original  and recovered text before and after the use of LLM, along with the similarity score, can be found at {\footnotesize  \url{https://tinyurl.com/OFSemWatTextLLM}}.

%



\section{Conclusions}

In this paper,
we propose a robust watemarking method, named OF-SemWat, which embeds a semantic textual description of the image (or  the input prompt) inside the image itself. 
%
Experiments show that the method is extremely robust against a wide variety of image processing, and  the embedded text can be retrieved also after inpainting, permitting to unveil the semantic modification the image has undergone by analyzing the mismatch between image and text.

Future works will focus on
integrating automatic tools to reveal the semantic discrepancies between image and recovered text, 
e.g.,  exploiting graphs-based methods (like FineMatch \cite{hua2024finematch}).
Embedding richer descriptions, as well as better exploring  the tradeoff between robustness, payload and invisibility,
are also interesting directions for investigation.




\bibliographystyle{IEEEbib}
\bibliography{OFBreferences}

\end{document}